\documentstyle[prb,aps,twocolumn,floats,epsf]{revtex}
\def\geqap{\,\raise 2pt \hbox{$>\kern-11pt \lower 5pt \hbox{$\sim$}$}\,}
\def\leqap{\,\raise 2pt \hbox{$<\kern-10pt \lower 5pt \hbox{$\sim$}$}\,}
\begin{document}
\draft
\twocolumn[\hsize\textwidth\columnwidth\hsize\csname @twocolumnfalse\endcsname
\title{Orbital Wave and its Observation in Orbital Ordered Titanates and Vanadates}
\author{Sumio Ishihara} 
\address{Department of Physics, Tohoku University, Sendai 980-8578, Japan}
\date{\today}
\maketitle
\begin{abstract}
We present a theory of the collective orbital excitation termed orbital wave   
in perovskite titanates and vanadates with the triply degenerate $t_{2g}$ orbitals. 
The dispersion relations of the orbital waves 
for the orbital ordered LaVO$_3$, YVO$_3$ and YTiO$_3$ 
are examined in the effective spin-orbital coupled Hamiltonians associated 
with the Jahn-Teller type couplings. 
We propose possible scattering processes for the Raman  
and inelastic neutron scatterings from the orbital wave and 
calculate the scattering spectra for titanates and vanadates. 
It is found that both the excitation spectra and the observation methods of the orbital wave 
are distinct qualitatively from those for the $e_g$ orbital ordered systems. 
\end{abstract}
]
\narrowtext
%
%
\section{introdcution}
\label{sec:intro}
When degenerate electron orbitals are partially filled in correlated 
electron systems, 
this is recognized to be an internal degree of freedom belonging to 
an electron as well as spin and charge.
This orbital degree of freedom has recently attracted much attention 
especially in transition-metal (TM) oxides.\cite{tokura,tokunaga}
In particular, the orbital ordering (OO) and fluctuation 
play a key role in anisotropic electric, magnetic and optical properties 
in correlated oxides. 

The collective orbital excitation in the orbital ordered state 
is known to be orbital wave (OW), 
in analogous to the spin wave in magnetically ordered state, 
and its quantized object is termed orbiton. 
The theoretical study of OW has started in the 
idealized spin-orbital coupled model where the continuous symmetry 
exists in the orbital space, \cite{cyrot} 
and progressed in the anisotropic orbital 
model.\cite{komarov}
A realistic calculation has been done 
in LaMnO$_3$ with the doubly degenerate $e_g$ orbitals. \cite{ishihara_eh}
Recently new peak structures observed in the Raman spectra in 
LaMnO$_3$ are interpreted as scatterings from OW.\cite{saitoh} 
Although there are some debates about origin of the newly found peaks, 
OW or multiphonons,\cite{gruninger,saitoh_r} 
the energy and polarization dependences of the observed Raman spectra  
show a good agreement with the calculation based on the OW interpretation. 

The perovskite titanates $R$TiO$_3$ and vanadates $R$VO$_3$ ($R$: rare-earth ion) 
with the triply degenerate $t_{2g}$ orbitals 
are another class of materials 
where OW is expected. 
One of the well known orbital ordered materials 
is YTiO$_3$ where the nominal valence of the titanium ion is $3+$ 
and one electron occupies the triply degenerate $t_{2g}$ orbitals. 
The orbital ordered state associated with the Jahn-Teller (JT) type 
lattice distortion has been confirmed by 
the resonant x-ray scatterings, NMR  
and the polarized neutron scattering experiments and so on. \cite{nakao,itoh,ichikawa,maclean}
These results almost coincide with each other: 
there are four different orbitals in a unit cell where 
the wave functions are given as  
$\frac{1}{\sqrt{2}}(d_{xy}+d_{yz})$, 
$\frac{1}{\sqrt{2}}(d_{xy}-d_{yz})$, 
$\frac{1}{\sqrt{2}}(d_{xy}+d_{zx})$ and 
$\frac{1}{\sqrt{2}}(d_{xy}-d_{zx})$. \cite{ishihara_ti}
This type of OO, termed the 
$(d_{y(x+z)}/d_{y(x-z)}/d_{x(y+z)}/d_{x(y-z)})$-type from now on, 
is also supported by previous calculations.\cite{mizokawa,sawada,mochizuki} 
Although an exotic orbital state is proposed recently by taking into account the 
quantum fluctuation, \cite{ulrich,khaliullin}
the predicted orbital state 
being incompatible with the crystal lattice symmetry
is different from the above type of OO. 

A series of $R$VO$_3$ is systematically examined in the recent studies.\cite
{mizokawa,sawada,ren,miyasaka,miyasaka2,bordet,kawano,noguchi,blake,ulrich_v,khaliullin_v,motome,fang,silva} 
Two electron occupy the triply degenerate orbitals in a V$^{3+}$ ions. 
A sequential phase transition is found in YVO$_3$; \cite{ren,miyasaka,miyasaka2}
the G-type OO (O-G) occurs at $T_{OO1}$=200K and the 
C-type antiferromagnetic (AFM) ordering (S-C) appears at $T_{N1}$=115K. 
With further decreasing temperature, 
another orbital and magnetic transitions appear at $T_{N2}=T_{OO2}$=77K where 
the C-type OO associated with the G-type AFM order
(the (S-G/O-C) order) is realized.  
On the other hand, in LaVO$_3$, 
the C-type AFM ordering occurs at T$_N$(=143K) and, at slightly below this 
temperature, the G-type OO (the (S-C/O-G) order) appears. \cite{miyasaka,miyasaka2,bordet}
Types of OO in vanadates are  
determined that the $d_{xy}$ orbital is occupied at all the 
vanadium sites and the $d_{yz}$ and $d_{zx}$ orbitals are alternately ordered 
in the $xy$ plane (the C-type OO), and in all direction (the G-type OO). \cite{noguchi,blake} 
These kinds of OO are termed the pure OO states, and the OO such as realized in 
YTiO$_3$ is termed the mixed OO from now on. 
It is suggested that the experimentally observed type of OO in vanadates 
associated with the JT type distortion 
explains the several optical and magnetic properties. \cite{motome,fang}

Here, we present a theory of OW 
in titanates and vanadates where the orbital ordered 
states are confirmed experimentally well. 
The dispersion relations of OW are examined  
in the (S-C/O-G)- and (S-G/O-C)-phases for LaVO$_3$, YVO$_3$ 
(the low temperature phase), respectively, 
and the $(d_{y(x+z)}/d_{y(x-z)}/d_{x(y+z)}/d_{x(y-z)})$-type OO for YTiO$_3$. 
The calculations are based on the effective spin-orbital coupled Hamiltonians associated 
with the JT type electron-lattice coupling. 
We propose possible scattering processes for the Raman  
and inelastic neutron scatterings from OW and 
calculate the spectra. 
It is found that the excitation spectra and the observation methods of OW 
are distinct qualitatively from those for the $e_g$ orbital ordered systems such as manganites. 

In Sec.~II, the model Hamiltonian for titanates and vanadates 
with the triply degenerate $t_{2g}$ orbitals are introduced. 
In Sec.~III, the dispersion relations of OW 
for LaVO$_3$, YVO$_3$ and YTiO$_3$ are examined. 
The scattering spectra for the Raman and inelastic neutron scatterings 
from OW are shown in Sec.~IV. 
Section V is devoted to summary and discussion. 

\section{Model Hamiltonian}
\label{sec:model}
The collective orbital excitations 
in the orbital ordered state 
are studied in the spin-orbital model derived from the 
generalized Hubbard Hamiltonian with the triply degenerate 
$t_{2g}$ orbitals: 
\begin{eqnarray}
{\cal H}
&=&
\varepsilon_d\sum_{i, \sigma, \gamma}d_{i \gamma \sigma}^\dagger d_{i \gamma \sigma}
+{\cal H}_{el-el} 
\nonumber \\
&+&
\sum_{\langle i j \rangle, \gamma, \gamma', \sigma} \left (t_{ij}^{\gamma \gamma'}
d_{i \gamma \sigma}^\dagger d_{j \gamma' \sigma} +H.c. \right ) , 
\label{eq:original}
\end{eqnarray}
with the electron-electron interaction term 
\begin{eqnarray}
{\cal H}_{el-el}&=&U\sum_{i, \gamma} n_{i \gamma \uparrow} n_{i \gamma \downarrow}
+U'\sum_{i, \gamma > \gamma'} n_{i \gamma} n_{i \gamma'}
\nonumber \\
&+& I\sum_{i, \gamma > \gamma', \sigma, \sigma'} 
d_{i \gamma \sigma}^\dagger d_{i \gamma' \sigma'}^\dagger
d_{i \gamma \sigma'}        d_{i \gamma' \sigma} 
\nonumber \\
&+& J \sum_{i, \gamma \ne \gamma'} 
d_{i \gamma  \uparrow}^\dagger d_{i \gamma  \downarrow}^\dagger
d_{i \gamma' \downarrow}  d_{i \gamma' \uparrow}         . 
\label{eq:humorig}
\end{eqnarray}
$d_{i \gamma \sigma}$ is the annihilation operator for 
the $t_{2g}$ electron at site $i$ with spin $\sigma=(\uparrow, \downarrow)$ 
and orbital $\gamma=(yz,zx,xy)$. 
The number operators are defined by 
$n_{i \gamma \sigma}=d_{i \gamma \sigma}^\dagger d_{i \gamma \sigma}$ 
and $n_{i \gamma}=\sum_\sigma n_{i \gamma \sigma}$. 
$U$ and $U'$ are the intra- and inter-orbital Coulomb interactions, 
respectively, $I$ is the exchange interaction,  and $J$ is the pair-hopping interaction. 
$t_{ij}^{\gamma \gamma'}$ in Eq.~(\ref{eq:original}) is the electron transfer integral 
between site $i$ with orbital $\gamma$ and 
nearest neighboring (NN) site $j$ with $\gamma'$. 
In the ideal perovskite structure, 
the hopping integral is simplified as  
$t_{ij}^{\gamma \gamma'}=
t_0 \delta_{\gamma \gamma'} (\delta_{\gamma=(lh)}+\delta_{\gamma=(kl)}) $ 
where $l$ indicates a direction of a bond connecting sites $i$ and $j$,  
and $(h,k,l)=(x,y,z),\ (y,z,x),\ (z,x,y)$. 
The electron hopping occurs through the O $2p$ orbitals in between the NN TM sites.

We derive the effective Hamiltonian 
for titanates where the nominal electron configuration of the TM ion is $d^1$. 
This Hamiltonian is defined in the Hilbert space where 
the electron number per site is restricted to be one or less 
due to the strong on-site Coulomb interactions. 
The Hamiltonian is classified by the intermediate states 
of the perturbational processes as derived in Ref.~\onlinecite{ishihara_ti}; 
\begin{eqnarray}
{\cal H}_{\rm Ti}={\cal H}_{T_1}+{\cal H}_{T_2}+{\cal H}_{E}+{\cal H}_{A_1}, 
\label{eq:eff}
\end{eqnarray}
with 
\begin{equation}
{\cal H}_{T_1}=-J_{T_1} \sum_{\langle ij \rangle}
\left ( {3 \over 4}+\vec S_i \cdot \vec S_j \right )  
\left ( A_-^l+B^l -C_+^l   \right )  , 
\label{eq:ht1}
\end{equation}
\begin{equation}
{\cal H}_{T_2}=-J_{T_2} \sum_{\langle ij \rangle}
\left ( {1 \over 4}-\vec S_i \cdot \vec S_j \right )  
\left ( A_-^l+B^l +C_+^l  \right )  , 
\label{eq:ht2}
\end{equation} 
\begin{equation}
{\cal H}_{E}=-J_{E} \sum_{\langle ij \rangle}
\left ( {1 \over 4}-\vec S_i \cdot \vec S_j \right )  
\frac{2}{3} \left (2A_+^l- C_-^l \right )     , 
\label{eq:he}
\end{equation}
\begin{equation}
{\cal H}_{A_1}=-J_{A_1} \sum_{\langle ij \rangle}
\left ( {1 \over 4}-\vec S_i \cdot \vec S_j \right )  
\frac{2}{3} \left ( A_+^l + C_-^l \right )     .  
\label{eq:ha1}
\end{equation}
The superexchange interactions are given by  
$J_{\Gamma}=t_0^2/E_{\Gamma}^{(2)}$ $(\Gamma=T_1,\ T_2,\ E, A_1)$ 
where $E_\Gamma^{(2)}$'s are the energies of the intermediate 
states: $E_{T_1}^{(2)}=U'-I$, $E_{E}^{(2)}=U-I$, $E_{T_2}^{(2)}=U'+I$ 
and $E_{A_1}^{(2)}=U+2I$ where 
the relations $U=U'+2I$ and $I=J$ 
are used. 
$J_{T_1}$ is the largest among them. 
The spin degree of freedom is described 
by the operator 
$\vec S_i=\frac{1}{2} \sum_{\gamma s s'} 
d_{i \gamma s}^\dagger \vec \sigma_{ss'} d_{i \gamma s'}$ 
with the Pauli matrices $\vec \sigma$. 
The orbital degree of freedom is represented by the  
eight orbital operators $O_{i \Gamma \gamma}$ 
classified by the irreducible representations 
of the $\rm O_h$ group as 
$(\Gamma \gamma)=
(Eu)$, $(Ev)$, $(T_2x)$, $(T_2y)$, 
$(T_2z)$, $(T_1x)$, $(T_1y)$, $(T_1z)$. 
These operators are defined by the generators of the SU(3) Lie algebra, 
i.e. the 3$\times$3 Gell-Mann matrices $\lambda_m$ $(l=m \sim 8)$ as \cite{gellmann}
\begin{equation}
O_{i \Gamma \gamma}={-1 \over \sqrt{2}} \sum_{\sigma, \alpha, \beta} 
d_{i \alpha \sigma}^\dagger \lambda_{m \alpha \beta} d_{i \beta \sigma} , 
\label{eq:orbop}
\end{equation}
where $(\Gamma \gamma, m)=(Eu, 8)$, $(Ev, 3)$
$(T_2x, 6)$, $(T_2y, 4)$, $(T_2z, 1)$, 
$(T_1x, 7)$, $(T_1y, 5)$, $(T_1z, 2)$. 
The operators $O_{iE\gamma}$ and $O_{i T_2 \gamma}$ 
represent the diagonal and off-diagonal 
components of the electric quadrupole moments, respectively, 
and $O_{i T_1 \gamma}$ does the magnetic dipole ones, i.e. 
the orbital angular momentum. 
By utilizing the above orbital operators, 
the orbital parts of the Hamiltonian in 
Eqs.~(\ref{eq:ht1})-(\ref{eq:ha1}) are given as 
\begin{eqnarray}
A^l_{\pm}= W_i^l W_j^l
   \pm O^l_{i Ev} O^l_{j Ev} , 
\label{eq:aaa}
\end{eqnarray}
\begin{eqnarray}
B^l&=&V_i^lW_j^l+W_i^lV_j^l,   
\label{eq:ddd}
\end{eqnarray}
\begin{equation}
C^l_{\pm}=2 \left (O_{i T_2 l} O_{j T_2 l} \pm O_{i T_1 l}O_{j T_1 l} \right ) , 
\label{eq:ccc}
\end{equation}
with 
\begin{eqnarray}
W_i^l&=&\frac{2}{3}-\sqrt{\frac{2}{3}} O_{i Eu}^l, 
\\
V_i^l&=&\frac{1}{3}+\sqrt{\frac{2}{3}} O_{i Eu}^l . 
\end{eqnarray}
$O^l_{i E u}$ and $O^l_{i E v}$ are defined by 
\begin{equation}
\left ( 
   \begin{array}{@{\,} c @{\,}}
   O^l_{i Eu} \\ 
   O^l_{i Ev}  
\end{array}
\right ) 
 =
\left ( 
   \begin{array}{@{\,} cc @{\,}}
   \ \ \ \cos {2 \pi \over 3}m_l & \sin{2 \pi \over 3}m_l  \\
   -\sin {2 \pi \over 3}m_l & \cos{2 \pi \over 3}m_l 
   \end{array}
   \right ) 
\left ( 
   \begin{array}{@{\,} c @{\,}}
   O_{i Eu} \\ 
   O_{i Ev}  
\end{array}
\right )   ,
\end{equation}
with $m_l=(1,2,3)$ for 
a direction of the bond $l=(x,y,z)$. 
In analogy with the spin Hamiltonian, 
$A_\pm^l$ and $B^l$ correspond to the $S_{iz}S_{jz}$ term, 
and $C_+^l$ and $C_-^l$ to the 
$S_{i+}S_{j-}+S_{i-}S_{j+}$ and 
$S_{i+}S_{j+}+S_{i-}S_{j-}$ terms respectively. 
The latter term originates form the pair hopping processes 
which break the conservation of the total electron number at each orbital, 
e.g. $\sum_{i} \langle n_{i xy } \rangle$. 

The effective Hamiltonian for vanadates, 
where the nominal electron configuration is $d^2$, 
is derived in a similar way from Eq.~(\ref{eq:original}). 
The $d^2$ state is assumed to be the lowest 
$^3T_1$ state due to the Hund rule. 
The explicit form is given by 
\begin{eqnarray}
{\cal H}_{\rm V}={\cal H}_{A_1}+{\cal H}_{E}+{\cal H}_{T_1}+{\cal H}_{T_2}, 
\label{eq:eff2}
\end{eqnarray}
with 
\begin{eqnarray}
{\cal H}_{A_1}&=&-J_{A_1} \sum_{\langle ij \rangle}
\frac{1}{6} \left ( 2+\vec S_i \cdot \vec S_j  \right )
\nonumber \\
&\times &
\left ( -A_+^l+A_-^l+B^l-2C_+^l \right) , 
\label{eq:ha1v}
\end{eqnarray}
\begin{eqnarray}
{\cal H}_{E}&=&-J_{E} \sum_{\langle ij \rangle}
\frac{1}{6} \left ( 1-\vec S_i \cdot \vec S_j \right) 
\nonumber \\
&\times &\left ( -A_+^l+A_-^l+B^l+C_+^l \right) , 
\label{eq:he2v}
\end{eqnarray} 
\begin{equation}
{\cal H}_{T_1}=-J_{T_1} \sum_{\langle ij \rangle}
\frac{1}{4} \left ( 1-\vec S_i \cdot \vec S_j \right) 
\left (  A_+^l- C_-^l \right)    , 
\label{eq:ht12v}
\end{equation}
\begin{equation}
{\cal H}_{T_2}=-J_{T_2} \sum_{\langle ij \rangle}
\frac{1}{4} \left ( 1-\vec S_i \cdot \vec S_j \right) 
\left (  A^l + C_-^l \right) . 
\label{eq:ht22v}
\end{equation}
We introduce 
the spin operator $\vec S_i$ with a magnitude $S=1$, 
and the exchange parameters 
defined by $J_\Gamma=t_0^2/\Delta E_\Gamma$ where 
$\Delta E_\Gamma=E_\Gamma^{(3)}-E_{T_1}^{(2)}$
with 
$\Delta E_{A_1}=U'-I$, 
$\Delta E_{E}=\Delta E_{T_1}=U'+2I$ and 
$\Delta E_{T_2}=U'+4I$. 
It is noted that 
a similarity between ${\cal H}_{\rm Ti}$ and ${\cal H}_{\rm V}$ is attributed to the fact 
that, with respect to the orbital degree of freedom, 
the high spin $^3T_1$ state for the $d^2$ configuration 
in the hole picture is equivalent 
to the $d^1$ state in the electron picture. 
Similar types of the  spin-orbital coupled Hamiltonian in the 
triply degenerate $t_{2g}$ orbitals are also derived by 
several authors for the $d^1$ and $d^2$ systems \cite{mochizuki,khaliullin,khaliullin_v,silva,kugel,kikoin}  

In addition to the electronic Hamiltonian, 
the electron-lattice interactions are introduced. 
In the $t_{2g}$ orbital systems, 
there are two kinds of the JT type interactions; 
\begin{eqnarray}
{\cal H}_{JT}&=&g_E \sum_{i , \gamma=(u,v)}Q_{i E \gamma}O_{i E \gamma}
\nonumber \\
&+&g_{T_2} \sum_{i,  \gamma=(x,y,z) } Q_{i T_2 \gamma} O_{i T_2 \gamma} , 
\label{eq:jt}
\end{eqnarray}
where $g_{E}$ and $g_{T_2}$ are the coupling constants 
and $Q_{i E \gamma}$ and $Q_{i T_2 \gamma}$ are the normal modes in an 
O$_6$ octahedron with symmetries E$_g$ and T$_{2g}$, respectively. 
$Q_{E \gamma}$ directly modifies the TM-O bond lengths and 
$Q_{T_2 \gamma}$ modifies the O-TM-O bond angles. 

Energy parameter values have been numerically evaluated 
by several authors.  
The effective exchange parameters $J^S$ 
in the Heisenberg model are obtain 
from the spin-wave dispersion relations as 
$J^S_z$=5.5meV ($z$ axis) and $J^S_{xy}$=5.8meV ($xy$ plane) 
in the (S-G/O-C) phase in YVO$_3$, $J^S_z=2.2-4$meV, 
$J^S_{xy}=$2.6meV in the (S-C/O-G)
phase in YVO$_3$ (Ref.~\onlinecite{ulrich_v}), 
and $J_z^S=J^S_{xy}$=3meV in YTiO$_3$ (Ref.~\onlinecite{ulrich}). 
The effective exchange parameters for the orbital operators in LaVO$_3$ 
are also estimated as 
$J^O_{z}\equiv 4t_0^2/(U'-I)$=33meV and $J^O_{xy}$=2meV (Ref.~\onlinecite{motome}). 
The JT stabilization energy is obtained from the LDA+U method \cite{motome}
as $E_{JT}$=27meV which is comparable or larger than the exchange interactions. 
The relativistic spin-orbit interaction which is not taken into account 
in the present model is about 0.4meV being much smaller than both the 
exchange and JT energies. \cite{sawada} 
This is consistent with the experimental results 
in the magnetic x-ray scattering in YTiO$_3$; 
the angular momentum separately estimated from the spin momentum 
is found to be negligible small. \cite{itoh_m}
Thus, the Hamiltonian ${\cal H}_{\rm Ti(V)}+{\cal H}_{JT}$ 
introduced above is the minimal model for examination of OW.  

Here we mention the implications of the theoretical model 
for the observed spin/orbital orders.  
As we have shown in Ref.~\onlinecite{ishihara_ti}, in the mean field theory, 
the large orbital degeneracy 
remains in the ferromagnetic (FM) ground state in ${\cal H}_{\rm Ti}$. 
A small perturbation, such as the Jahn-Teller type distortion, the relativistic 
spin-orbit coupling, the GdFeO$_3$-type lattice distortion, 
lifts the degeneracy. 
We suppose that the Jahn-Teller type distortion with the $T_{2g}$ symmetry, $g_{T_2}Q_{iT_2}$ 
in Eq.~(\ref{eq:jt}), plays a key role to stabilize the observed mixed orbital order. 
The (S-C/O-G) order for LaVO$_3$ is 
reproduced by ${\cal H}_{\rm V}$. 
The FM (AFM) order along the $c$ axis (in the $ab$ plane) 
is attributed to the alternate (uniform) alignment of 
the $d_{yz}$ and $d_{zx}$ orbitals (the $d_{xy}$ orbital) along the $c$ axis (in the $ab$ plane). 
The (S-G/O-C) order for the YVO$_3$ 
is obtained by ${\cal H}_{\rm V}$ and the Jahn-Teller type interaction with the $E_g$ symmetry. 
The AFM spin order and the uniform 
alignment of the $d_{yz} (d_{zx})$ orbital along the $c$ axis is stabilized 
cooperatively, as discussed in the next section. 

\section{Orbital Wave}
The dispersion relations of OW in the orbital ordered states 
are obtained by utilizing the Holstein-Primakoff transformation 
for the generators in the SU(3) algebra. \cite{janssen,klein}
For example, at a site where the $d_{xy}$ orbital is occupied, 
there are two excitation modes; 
an excitation between the $d_{xy}$ and $d_{yz}$ orbitals denoted by a boson 
operator $y$ ($y^\dagger$), 
and that between $d_{xy}$ and $d_{zx}$ denoted by $x$ ($x^\dagger$). 
The orbital operators $O_{i \Gamma \gamma}$ are transformed 
into these boson operators as 
\begin{eqnarray}
O_{i E u}&=&\sqrt{\frac{2}{3}}-\sqrt{\frac{3}{2}}\left ( n_{ix}+n_{iy} \right ) , 
\nonumber \\
O_{i E v}&=&\frac{1}{\sqrt{2}} \left ( n_{iy}-n_{ix} \right ) , 
\nonumber \\
O_{i T_{\alpha } z}&=& {-i \choose 1}_\alpha 
\left (y_i^\dagger x_i \pm x_i^\dagger y_i \right ) , 
\nonumber \\
O_{i T_{ \alpha } y }&=& {i \choose 1}_\alpha
\left (\sqrt{1-N_i}y_i \pm y_i^\dagger \sqrt{1-N_i} \right ) , 
\nonumber \\
O_{i T_{ \alpha} x}&=& {-i \choose 1}_\alpha
\left (x_i^\dagger \sqrt{1-N_i} \pm \sqrt{1-N_i} x_i \right ) , 
\label{eq:hp}
\end{eqnarray}
with $\alpha=(1,2)$ . 
The plus and minus signs in $O_{i T_{\alpha }x}$, 
$O_{i T_{\alpha}y }$ and $O_{i T_{\alpha}z}$ are for the $\alpha=1$ and $\alpha=2$ cases, 
respectively.  
We define $N_i=n_{ix}+n_{iy}$ with $n_{i x}=x_i^\dagger x_i$ and  $n_{i y}=y_i^\dagger y_i$. 
In the linear spin wave approximation, 
$\sqrt{1-N_i}$ is replaced by 1.  
We have cheked that the Green's function method for the operators $O_{i \Gamma \gamma}$, i.e. 
$G_{\Gamma \gamma \Gamma' \gamma'}(t-t', \vec r_i-\vec r_{i'})=
-i\theta(t-t')\langle [ O_{i \Gamma \gamma}(t) , O_{i' \Gamma' \gamma'}(t') ] \rangle$, 
with the decoupling approximation, 
reproduces the calculated results for OW with the boson method introduced above. 

\begin{figure}
\epsfxsize=0.8\columnwidth
\centerline{\epsffile{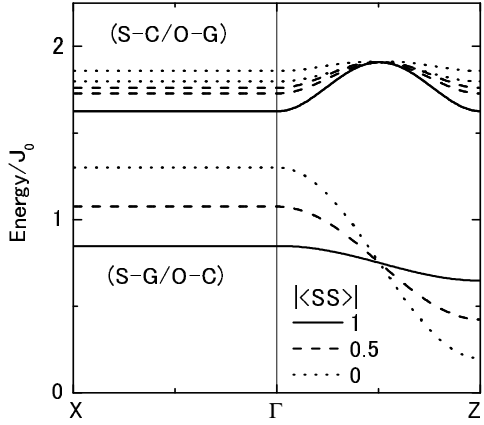}}
\caption{Dispersion relations of OW in the (S-C/O-G) and 
(S-G/O-C)-phases for LaVO$_3$ and YVO$_3$ (the low temperature phase), 
respectively. 
The absolute values of the 
spin correlation function between NN sites 
$| \langle \vec S_i \cdot \vec S_{i+\delta} \rangle |$ 
are 1 (bold lines), 0.5 (broken lines) and 0 (dotted lines). 
The parameter values are chosen to be $I/U'$=0.125. 
$g_EQ_E/J_0$=0.8 and $g_{T_2}Q_{T_2}$=0. 
The dispersion curves in the (S-G/O-C) phase and the curve in the (S-C/O-G) phase 
with  $|\langle \vec S_i \cdot \vec S_{i+\delta} \rangle|=1$ are doubly degenerate. 
}
\label{fig1}
\end{figure}
In Fig.~1, we present the dispersion relations of OW 
in the (S-C/O-G) and (S-G/O-C) phases 
for LaVO$_3$ and YVO$_3$ (the low temperature phase), respectively. 
The parameter values are chosen to be $I/U' \equiv R$=0.125, $g_E Q_E/J_0$=0.8 and 
$g_{T_2}Q_{T_2}$=0. 
The energy parameters are normalized by $J_0=4t_0^2/(U'-I)$ which is  
estimated to be about 33meV for LaVO$_3$ (Ref.~\onlinecite{motome}) and 
is supposed to be smaller in YTiO$_3$ due to the larger GdFeO$_3$-type 
lattice distortion. 
The ratios of the exchange parameters are represented by the parameter $R$ as 
$J_{T_2}/J_{T_1}=J_E/J_{T_1}=(1-R)/(1+2R)$
and $J_{A_1}/J_{T_1}=(1-R)/(1+4R)$. 
Both the (S-C/O-G)- and (S-G/O-C)-phases, there are four modes of OW attributed to the two 
different orbital occupied sites in a unit cell;  
there are the excitations $y_A$ ($d_{xy} \rightarrow d_{yz}$) and 
$z_A$ ($d_{zx} \rightarrow d_{yz} $) 
at site A where the $d_{xy}$ and $d_{zx}$ orbitals are occupied, and 
the excitations $x_B$ ($d_{xy} \rightarrow d_{zx}$) and 
$z_B$ ($d_{yz} \rightarrow d_{zx} $) 
at site B where the $d_{xy}$ and $d_{yz}$ orbitals are occupied.  
Two of the four, i.e. 
$y_A$ and $x_B$,  are the local modes within the linear spin wave theory 
and do not show dispersions. 
This character does not depend on the spin arrangements. 
These local modes originate from the facts that (1) the excited $d_{xy}$ hole does not  
hop along the $z$ axis due to the orbital symmetry, and (2) 
a coherent motion of the excited $d_{xy}$ hole in the $xy$ plane 
are impossible, since this motion is associated with increasing the number of the 
wrong orbital arrangements. 
The latter implies that the orbital exchange processes do not recover the 
wrong orbital arrangements, and the triply degenerate orbital model is 
qualitatively different from the Heisenberg model with $S=1$. 
The remaining modes, $z_A$ and $z_B$, are dispersive along the $z$ direction. 
The dispersion relation of 
OW in the (S-C/O-G) phase is explicitly obtained as  
$E(\vec k)=\frac{8}{6}J_{A_1} \sqrt{ (K_2^x+K_2^z)^2 -(\frac{3}{2}K_1^z \cos ak_z)^2 }$ 
where we assume $I=0$ and $g_{E}Q_E=g_{T_2}Q_{T_2}=0$. 
We introduce $K_2^{x(z)}=2+\langle \vec S_i \cdot \vec S_{i+\delta_x(\delta_z)} \rangle$  
and $K_1^{x(z)}=1-\langle \vec S_i \cdot \vec S_{i+\delta_x(\delta_z)} \rangle$. 
This energy has its minimum at $k_z=0$ and 
the energy gap is attributed to the anisotropy in the orbital space, i.e. 
a lack of the SU(3) symmetry, in the orbital part of the Hamiltonian. 
In comparison with the OW in the (S-C/O-G) phase, 
the OW in the (S-G/O-C) phase is barely stable; 
with decreasing the spin correlation which corresponds to 
increasing temperature toward $T_N$, 
a remarkable softening around $\vec k=(0,0,\pi)$ occurs.
In the case where $I=0$ and $g_{T_2}Q_{T_2}=0$, 
the dispersion relation is given as  
$E(\vec k)=\frac{2}{3} J_{A_1}(2K_2^z-K_1^z) \cos ak_z +\sqrt{\frac{3}{2}}g_{E}Q_{E}$ 
which has its minimum value at $k_z=\pi$, 
and the energy gap is attributed to the JT type interaction. 
This result suggests an instability of the (S-G/O-C) phase to the (S-C/O-G) one  
with increasing temperature. 
This is consistent with the experimental fact that in $R$VO$_3$ the O-C phase 
appears associated with the S-G order, and 
is changed into the (S-C/O-G) phase at 77K in YVO$_3$.\cite{miyasaka2} 
\begin{figure}
\epsfxsize=0.9\columnwidth
\centerline{\epsffile{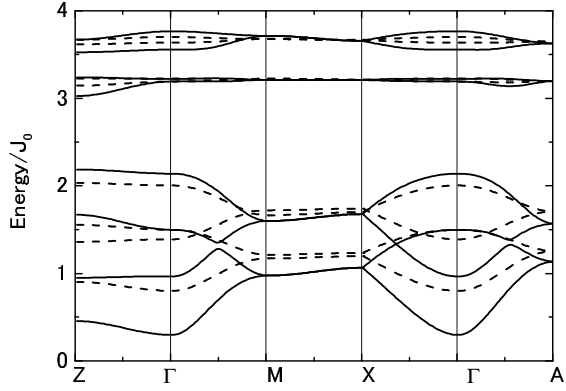}} 
\caption{Dispersion relations of OW 
in the $(d_{y(x+z)}/d_{y(x-z)}/d_{x(y+z)}/d_{x(y-z)})$-type 
OO for YTiO$_3$.
The bold and broken lines are for the FM and paramagnetic 
states, respectively. 
The parameter values are chosen to be 
$I/U'$=0.125, $g_E Q_E/J_0$=$g_{T_2}Q_{T_2}/J_0$=1.2. 
The Brillouin zone for the tetragonal symmetry is adopted. }
\label{fig2}
\end{figure}

In Fig.~2, 
we present the dispersion relations of OW 
in the $(d_{y(x+z)}/d_{y(x-z)}/d_{x(y+z)}/d_{x(y-z)})$-type OO for YTiO$_3$. 
The FM and paramagnetic spin orders are assumed, 
and the exchange parameter is taken to be $R\equiv I/U'=0.125$. 
The JT type interaction parameters are chosen to be 
$g_EQ_E/J_0=g_{T_2}Q_{T_2}/J_0=1.2$, 
where $g_E Q_E$ is larger than that in vanadates 
and $g_{T_2}Q_{T_2}$ is introduced.
This is based on a consideration that, 
in comparison with vanadates, 
$J_0$ is supposed to be smaller 
due to the large GeFeO$_3$-type lattice distortion, and 
$Q_{T_2}$ is found to be larger in the crystal structure of YTiO$_3$. \cite{maclean,bordet,kawano}
In contrast to the case of vanadates, 
all the eights modes, attributed to the four different orbitals in a unit cell,  
are dispersive along all the directions in the Brillouin zone. 
This originates from the OO states with the mixed orbitals where 
the excitations propagate along the three directions. 
The OW dispersions are classified into the two groups: 
for example, in the $\frac{1}{\sqrt{2}}(d_{xy}+d_{yz})$ orbital occupied site, 
the higher energy bands with smaller band widths   
are mainly attributed to the excitations  
to the $d_{zx}$ orbital, and 
the lower energy ones with a larger width 
are attributed to the excitations to the 
$\frac{1}{\sqrt{2}}(d_{xy}-d_{yz})$ orbital. 

\section{observation of orbital wave}
\label{sec:observation}
\subsection{raman scattering}
\begin{figure}
\epsfxsize=0.8\columnwidth
\centerline{\epsffile{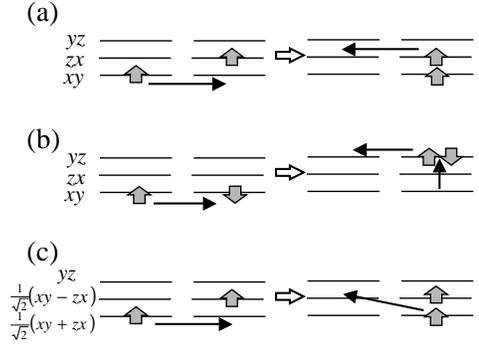}}
\caption{Scattering processes in the Raman scattering. (a) and (b) 
The two-orbiton scattering processes 
in the pure OO states. 
(c) The one-orbiton scattering processes in the mixed OO state.  
}
\label{fig3}
\end{figure}

As introduced in Sec.~I, 
in orbital ordered LaMnO$_3$, 
the new peak structures in the Raman spectra  
were explained successfully as the scattering from OW. \cite{saitoh}
Here we consider the Raman scattering as 
a possible probe to detect the OW in titanates and vanadates 
with the triply degenerate $t_{2g}$ orbitals. 
It is considered the inter-site scattering processes where 
OW's are excited through the electronic exchange 
processes between the NN TM sites. \cite{inoue,okamoto_ow}
This is attributed to the fact that 
the lowest electronic excitations in titanates and vanadates occur  
across the Mott-Hubbard gap, unlike the manganites 
where the electronic excitations across the charge-transfer gap dominate 
the excitation of OW. 
Depending on the types of OO, 
there are the following two scattering processes: 
(i) The two-orbiton scattering: 
Consider a pair of the NN TM sites where 
the occupied orbitals are different and the pure orbitals (Fig.~\ref{fig3}), such 
as the OO in vanadates. 
Through the second order processes of the interaction between photons and electrons, 
electrons at the two sites are exchanged and, at the final state, 
the occupied orbitals are changed at both the sites. 
This is the analogous to the two magnon Raman scattering in the antiferromagnets. 
Another two-orbiton process occurs in a pair of the NN TM site 
where the same kind of orbitals are occupied. 
In the intermediate state of the scattering process, where 
two electrons occupy the same orbital at a site, 
the occupied orbital is changed 
due to the pair-hopping interaction. 
At the final state, the two orbitons are created.  
(ii) The one-orbiton scattering: 
When the electrons occupy the so-called mixed orbital such as that in YTiO$_3$, 
the electron hops from one orbital ($\gamma$) to the 
different orbital ($\gamma'$) in its NN TM site. 
When this electron comes back to the orbital $\gamma'$ in the initial site, 
one orbiton is excited. 

There are alternate two scattering processes from OW 
where one orbton is created at a TM site. 
(i) An electron is excited from a orbital $\gamma$ to 
$\gamma'$ at the same site associated with 
a creation of odd-parity phonons. 
Then, these phonons are annihilated by emitting a photon. 
Such kinds of the Frank-Condon processes have been considered 
for the orbiton+phonon excitation in the optical conductivity spectra, 
and for the multiphonon excitation in the Raman spectra.\cite{allen} 
The total scattering-cross section ratio of this process to the 
inter-site process is of the order of $ 10^{-3} \sim 10^{-2}$. 
(ii) The incident photon excites an electron from the $3d$ $\gamma$ orbital to 
one of the $4p$ orbitals at the same site. 
Then, this electron is relaxed to the $3d$ $\gamma'$ orbital by emitting a photon. 
This scattering-cross section ratio to the inter-site process is 
estimated for titanates and vanadates to be of the order of $10^{-1}$. 

Here, we calculate the Raman spectra for the inter-site scattering process. 
It is supposed that this process provides the main contribution 
for the OW scattering, in particular, in the two-orbiton energy regions. 
The energy, momentum and polarization of initial (scattered) photon are 
$\omega_{i}(\omega_f)$, $\vec k_{i}(\vec k_f)$ and $\lambda_{i}(\lambda_f)$, respectively. 
The differential scattering cross section from OW is given as 
\begin{eqnarray}
\frac{d^2 \sigma}{ d\Omega dE_f}&=&
\sigma_T  \frac{\omega_f}{\omega_i} 
\left ( \frac{ma^2}{\hbar^2} \right)^2
\frac{1}{2 \pi \hbar} \int dt e^{i \hbar \left ( \omega_{f}-\omega_i \right ) t} 
\nonumber \\
&\times& 
\sum_{ll'} P_{ll'} S^{ll'}(t) , 
\end{eqnarray}
with $\sigma_T=(e^2/mc^2)^2$ and a bond length $a$. 
$P_{ll'}$ is the polarization factor given by 
\begin{eqnarray}
P_{ll'}=
\left ( \vec e_{k_i \lambda_i} \right )_l \left ( \vec e_{k_i \lambda_i} \right )_{l'}
\left ( \vec e_{k_f \lambda_f} \right )_l \left ( \vec e_{k_f \lambda_f} \right )_{l'} , 
\end{eqnarray}
and $S^{ll'}(t)$ is the dynamical correlation function defined by 
\begin{eqnarray}
S^{ll'}(t)= 
\langle  
{\cal K}^{l}(t) 
{\cal K}^{l'}  (0) \rangle , 
\end{eqnarray}
with  
\begin{equation}
{\cal K}^l=\sum_\Gamma \sum_i \sum_{\delta_l} {\cal H}_{\Gamma}(i,i+\delta_l) . 
\end{equation}
${\cal H}_{\Gamma}(i, i+\delta_l)$ is 
a term of the effective Hamitonian given in 
Eqs.~(\ref{eq:eff}) and (\ref{eq:eff2}) 
concerning with a bond connecting site $i$ and site $i+\delta_l$. 
Here, $l(=x,y,z)$ indicates a 
direction of the bond and $\delta_l=\pm \hat l$. 
The index $\Gamma(=T_1, T_2, A_1, E)$ classifies the intermediate states. 
The exchange interaction $J_\Gamma=t_0^2/\Delta E_\Gamma$ in ${\cal H}_\Gamma$ is 
replaced by $t_0^2/(\Delta E_\Gamma-(\omega_{i}-\omega_f ))$ 
in ${\cal H}_\Gamma(i, i+\delta_l)$. 
This expression is obtained from the second order processes of the interactions 
between photons and electronic currents between the NN TM sites. 

\begin{figure}
\epsfxsize=0.85\columnwidth
\centerline{\epsffile{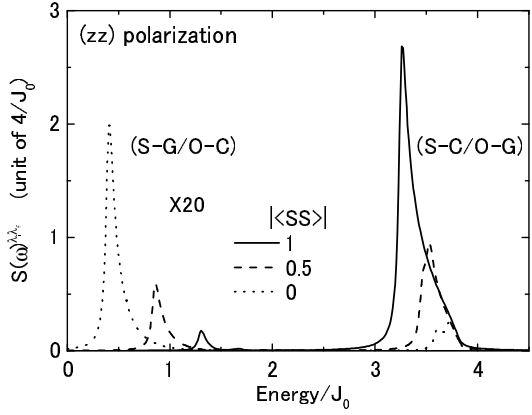}}
\caption{
Raman spectra $S^{\lambda_i \lambda_f}
(\omega)$ for the (S-C/O-G)- and (S-G/O-C) phase 
for LaVO$_3$ and YVO$_3$ (the low temperature phase), respectively. 
The two-orbiton scattering is considered. 
The absolute values of 
the spin correlation function between NN sites 
$|\langle \vec S_i \cdot \vec S_{i+\delta} \rangle |$ 
are 1 (bold lines), 0.5 (broken lines) and 0 (dotted lines). 
Both the incident and scattered photon 
polarizations are 
parallel to the $z$ axis. 
Other parameter values are the same as those in Fig.~1. 
}
\label{fig4}
\end{figure}
We introduce the orbiton operators ${\widetilde \psi(\vec k)}$ 
with the energies $E(\vec k)$ 
which are obtained by diagonalizing the Hamiltonian 
represented by the Holstein-Primakoff boson operators $\psi(\vec k)$. 
This Bogoliubov transformation is given by 
the matrix $V(\vec k)$ as 
\begin{equation}
\widetilde{\psi}_{\alpha}(\vec k)=\sum_\beta \psi_{\beta}(\vec k) V _{\beta \alpha}(\vec k)  . 
\end{equation}
${\widetilde \psi}(\vec k)$ has $2N$ components 
in the system where the number of the OW mode is $N$, 
and ${\widetilde \psi}_\alpha(\vec k)$ $(\alpha > N)$ 
is the creation operator with the condition 
${\widetilde \psi}_\alpha(\vec k)={\widetilde \psi}_{\alpha-N}(-\vec k)^\dagger$. 
For example, 
\begin{eqnarray}
{\widetilde \psi}(\vec k)=\left \{a_{1}(\vec k), \cdots a_{N}(\vec k), 
a_{1}^\dagger(-\vec k), \cdots a_{N}^\dagger(-\vec k) \right \} . 
\end{eqnarray}
By using the oprators, 
we obtain 
\begin{eqnarray}
{\cal K}^l&=&\sum_{\vec k, \alpha, \beta} {\widetilde \psi}_\alpha (\vec k)^\dagger 
 h^l_{\alpha \beta} (\vec k)
{\widetilde \psi}_{\beta}(\vec k)
\nonumber \\ 
&+& \frac{1}{2} \sum_\alpha 
\left ( 
g_{\alpha}^l {\widetilde \psi}_\alpha (0) +
g_{\alpha}^{l \ast} {\widetilde \psi}^\dagger_\alpha (0) \right ) , 
\label{eq:k2}
\end{eqnarray}
where $h_{\alpha \beta}^l(\vec k)$ and $g_\alpha^l$ are the coefficients. 
In the pure OO states, the second term vanishes, i.e. the one-orbiton scattering is 
prohibited. 
Then
the Fourier transform of the dynamical correlation function in the two-orbiton Raman scattering 
is given by 
\begin{eqnarray}
S^{ll'}(\omega)&=&N \sum_{\vec k} \sum_{\alpha = N+1}^{2N} \sum_{\beta=1}^N
\delta \left \{ \omega-E_{\alpha}(\vec k)-E_{\beta}(\vec k) \right \}
\nonumber \\
&\times& 
\left \{
h^l_{ \alpha \beta}(\vec k) h^{l'}_{ \beta \alpha}(\vec k)+
h^l_{ \alpha \beta}(\vec k) h^{l'}_{ \alpha-N \beta+N}(\vec k) 
\right \} . 
\label{eq:two}
\end{eqnarray}
We neglect the orbiton-orbiton interaction for simplicity, and 
$S^{ll'}(\omega)$ is represented by the convolution of the two OW modes.  
On the other hand, in the one-orbiton scattering, 
the dynamical correlation function reflects the OW at the momentum $\vec k=0$ as 
\begin{eqnarray}
S^{ll'}(\omega)=4N \sum_{\alpha=1}^N  g_\alpha^l g_\alpha^{l'}
\delta \left \{ \omega-E_\alpha(0) \right \} . 
\label{eq:one}
\end{eqnarray}
\begin{figure}
\epsfxsize=0.7\columnwidth
\centerline{\epsffile{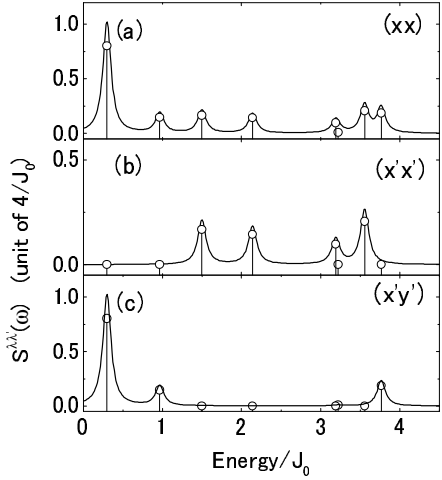}}
\caption{
Raman spectra for the 
$(d_{y(x+z)}/d_{y(x-z)}/d_{x(y+z)}/d_{x(y-z)})$-type 
OO state with FM order 
for YTiO$_3$.  
The one-orbiton scattering is considered. 
Symbols with vertical lines are the spectral weights of the 
delta-function.  
Continuous lines indicate $S^{\lambda_i \lambda_f}(\omega)$ 
where the delta-functions are replaced by the Lorentz functions and 
the width of the function is chosen to be $0.25J_0$. 
The polarizations of the incident and scattered photons 
are denoted as $(\lambda, \lambda')$.   
Other parameter values are the same as those in Fig.~2.  
}
\label{fig5}
\end{figure}
In Fig.~\ref{fig4}, 
the Raman spectra by OW in the (S-C/O-G) phase for LaVO$_3$, 
and in the (S-G/O-C) phase for YVO$_3$ (the low temperature phase) 
are presented. 
The two-orbiton scattering processes are considered. 
In spite of this processes, 
a sharp peak structure appears at the lower edge of the continuum. 
This is attributed to the one-dimensional character of the OW and 
the factor $ h^l_{\alpha \beta} (\vec k)$ in Eq.~(\ref{eq:k2})  
which enhances the lower edge. 
As for the selection rule, 
the Raman scattering is only active for the $(zz)$ polarization  
where both the incident and scattered light polarizations, 
$\vec e_{\vec k_i \lambda_i}$ and $\vec e_{\vec k_f \lambda_f}$, 
are parallel to the $z$ axis.  
Through the interaction with the $z$-polarized lights, 
two electrons are exchanged between the 
$d_{yz}$ ($d_{zx}$) orbitals in the NN sites along the $z$ axis. 
We mention that the local modes discussed in the previous section, 
i.e. the $d_{xy} \rightarrow d_{yz}$ and $d_{xy} \rightarrow d_{zx}$ excitations,  
are not detected by the Raman scattering, since the exchange processes do not 
occur between the $d_{xy}$ and $d_{yz}$ ($d_{zx}$) orbitals. 
In Fig.~\ref{fig5}, we show the Raman spectra from OW in the 
$(d_{y(x+z)}/d_{y(x-z)}/d_{x(y+z)}/d_{x(y-z)})$-type OO. 
The one-orbiton scattering is considered in the calculation. 
The spectra being active for the $(xx)$ and  $(x'x')$ polarizations 
are the $A_{1g}$ modes, and   
those for the $(xx)$ and $(x'y')$ ones are the $B_{1g}$ ones. 
Here, the $x$, $y$ and $z$ directions are chosen to be parallel to the TM-O bonds, 
and $x'=\frac{1}{\sqrt{2}}(x+y)$ and $y'=\frac{1}{\sqrt{2}}(-x+y)$. 
It is worth noting that all modes are inactive for the $(zz)$ polarization 
in contrast to the case of vanadates. 
This originates from a cancellation from the one-orbiton scattering contributions  
from site $i$ and its NN site $i+\delta_z$ along the $z$ direction 
where the occupied orbitals have the mirror symmetry in terms of the $xy$ plane 
between the two. 

\begin{figure}
\epsfxsize=0.7\columnwidth
\centerline{\epsffile{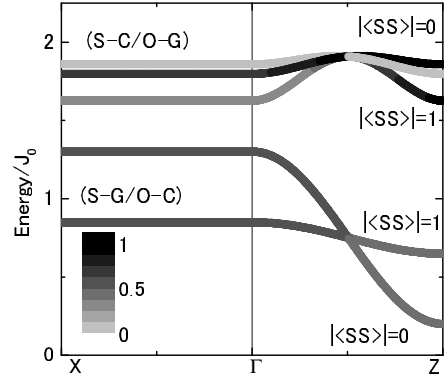}}
\caption{
Contour map of the scattering intensity $S^{zz}(\omega)$ in 
the inelastic neutron scattering in the (S-C/O-G) and (S-G/O-C)-phases 
for LaVO$_3$ and YVO$_3$ (the low temperature phase), respectively. 
The intensity is plotted as a unit of $2/J_0$. 
The spin polarizations of both the incident and 
scattered neutrons 
are chosen to be parallel to the $z$ axis. 
The reciprocal vector is $\vec G=(000)$. 
Other parameter values are the same as those in Fig.~1. 
}
\label{fig6}
\end{figure}
\subsection{inelastic neutron scattering}
Although the Raman scattering is a possible probe to detect OW as shown 
in the previous section, 
its observation is limited to be the OW at 
zero momentum in the one-orbiton scattering,  
and the joint density of states of OW 
in the two-orbiton scattering. 
In the $t_{2g}$ OO systems of our present interest, 
it is possible to detect by the inelastic neutron scattering. 
Formulate the differential scattering cross section 
in the scattering of initial (scattered) neutron 
with momentum $\vec k_i(k_f)$, energy $\omega_i (\omega_f)$, 
and polarization $l_i (l_f) \ (=x,y,z)$. 
The scattering cross section is given by 
\begin{eqnarray}
\frac{d \sigma^2}{d \Omega d \omega_f}&=&
\left ( \frac{\gamma e^2}{m_N c^2} \right )^2 
\left ( \frac{1}{2} gF(\vec K) \right )^2
\frac{k_f}{k_i}\nonumber \\
&\times&\sum_{l_i l_f} 
\left (\delta_{l_i l_f}-\kappa_{l_i} \kappa_{l_f} \right )
S^{l_i l_f}(\vec K, \omega), 
\end{eqnarray}
with $\vec K=\vec k_i-\vec k_f$, $\omega=\omega_i-\omega_f$, 
and $\vec \kappa=\vec K/|\vec K|$. 
$S^{l_i l_f}(\vec K , \omega)$ 
is the Fourier transform of the 
dynamical correlation function for the angular momentum operators $L_{i d}^l$
defined by 
\begin{eqnarray}
S^{l_i l_f}(\vec r_{id}-\vec r_{i'd'}, t)
&=&\langle L_{i d}^{l_i}(t) L_{i' d'}^{l_f}(0) 
\rangle , 
\nonumber \\
&=& 2\langle O_{i d T_1 l_i}(t)  O_{i' d' T_1 l_f}(0)   \rangle , 
\end{eqnarray}
where $\vec r_{id}$ is the position of 
the $d$-th TM ion in the $i$-the unit cell. 
In the linear spin wave approximation, 
$S(\vec K, \omega)$ is obtained as 
\begin{eqnarray}
S^{l_i l_f}(\vec K, \omega)&=&2N\sum_{d d'}
\sum_{\alpha =1}^N \delta \left \{ \omega-E_\alpha(\vec K) \right \}
\nonumber \\
&\times&
D^{l_i \ast}_{d \alpha} D^{l_f}_{d' \alpha}
e^{i \vec G \cdot \vec \delta_{dd'}} , 
\end{eqnarray}
where $D_{d \alpha}^l$ is defined by the Fourier transform 
of the angular momentum operator  
\begin{eqnarray}
O_{d T_{1} l} (\vec k)=\sum_{\alpha}
D_{d \alpha }^l(\vec k) {\widetilde \psi}_{\alpha}(\vec k) . 
\end{eqnarray}
$\vec \delta_{dd'}$ is a vector connecting the $d$- and $d'$-th TM ions 
in the same cell, and $\vec G$ is the reciprocal lattice vector. 
In contrast to the Raman scattering, 
the momentum dependence of the dispersion relation is detectable. 
The magnitude of the scattering intensity is expected to be the same order with 
that for the magnetic neutron scattering in magnets with $L=1$.

\begin{figure}
\epsfxsize=0.8\columnwidth
\centerline{\epsffile{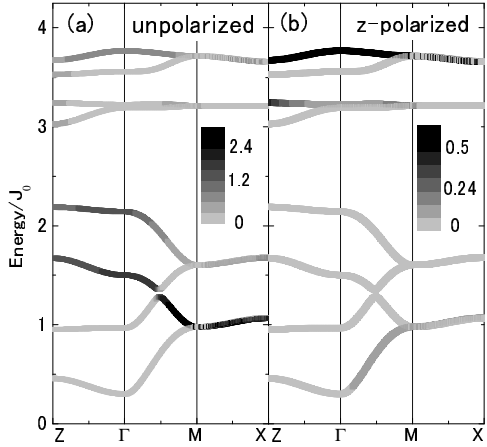}}
\caption{
Contour map of the inelastic neutron scattering spectra 
$S^{ll'}(\omega)$ for 
the $(d_{y(x+z)}/d_{y(x-z)}/d_{x(y+z)}/d_{x(y-z)})$-type OO state 
with FM order for YTiO$_3$.  
The intensity is plotted as a unit of $2/J_0$. 
The spin polarizations of the 
incident and scattered neutrons are chosen to be (a) unpolarized, and 
(b) parallel to the $z$ axis. 
The reciprocal vector is $\vec G=(000)$. 
Other parameter values are the same with those in Fig.~2. 
}
\label{fig7}
\end{figure}

We present the scattering intensities of the inelastic neutron scattering 
in the (S-C/O-G) and (S-G/O-C)-phases  (Fig.~\ref{fig6}), and 
those in the FM order with 
the $(d_{y(x+z)}/d_{y(x-z)}/d_{x(y+z)}/d_{x(y-z)})$-type OO (Fig.~\ref{fig7}). 
It is noted that 
the dispersive OW modes in vanadates are only detected by the $z$ polarized neutron;  
the angular momentum $L_z$ induces the excitations between the $d_{yz}$ and $d_{zx}$ orbitals. 
The local modes discussed in the previous section 
are active for the $x$ and $y$ polarized neutrons (not shown in Fig.~\ref{fig6}). 
Therefore, the modes of OW are identified by utilizing the polarized neutron scattering. 
This is also seen in the contour map of the scattering intensity 
in YTiO$_3$; 
the scattering intensity for the OW modes in the higher energy bands are 
remarkable in the $z$-polarized neutrons. 
This implies that these modes mainly consist of 
the excitations between the 
$d_{zx}$ and $d_{yz}$ orbitals as explained in Sec.~III. 

\section{summary and discussion}
In this paper, 
a theory of OW in perovskite titanates and vanadates 
with the triply degenerate $t_{2g}$ orbitals is present. 
We examine the dispersion relations of OW 
in the (S-C/O-G)- and (S-G/O-C) phases 
for LaVO$_3$ and YVO$_3$ (the low temperature phase), respectively and 
that in the $(d_{y(x+z)}/d_{y(x-z)}/d_{x(y+z)}/d_{x(y-z)})$-type OO for YTiO$_3$. 
We demonstrated that characteristics of OW in these systems can be detected 
by utilizing the Raman and inelastic neutron scatterings. 

In comparison with the OW in the manganites where 
the $e_g$ orbital is ordered, 
both the excitation spectra and the observation methods 
are distinct qualitatively in the present $t_{2g}$ orbital system. 
In particular, 
this is remarkably seen in vanadates where the 
so-called pure orbitals are ordered. 
Thus, the selection rules for the Raman and neutron scatterings are strict. 
For example, the two-orbiton scatterings with the $z$ polarized photons dominate the Raman spectra. 
This is attributed to the orthogonality 
of the electron transfer integral between the NN vanadium sites. 
In the actual vanadates, 
there is the GdFeO$_3$-type lattice distortion which may make 
the one-orbiton scatterings possible. 
In the recent Raman scattering experiments in $R$VO$_3$, 
a new peak appears around 60meV in the (S-C/O-G) phase.\cite{miyasaka3} 
It is confirmed that this peak is active in the $(zz)$ polarization configuration. 
We expect that this peak originates from OW excited  
by the two-orbiton scattering processes as discussed in Sec. III (Fig.~\ref{fig4}). 

In the case of YTiO$_3$,  
the dispersion relation of OW and 
the Raman and neutron scattering spectra are more complicated than those in vanadates.  
This is because of the mixed OO state with the four different orbitals in a unit cell. 
The present results are also distinct from those proposed in Ref.~\onlinecite{khaliullin};   
the orbital excitations are examined in the OO states with high symmetry being 
different from the $(d_{y(x+z)}/d_{y(x-z)}/d_{x(y+z)}/d_{x(y-z)})$-type OO 
and incompatible with the crystal symmetry of YTiO$_3$. 
In the present results, as shown in Fig.~2, 
it is found, in contrast to the previous results,\cite{khaliullin} that 
there are the anisotropy of the dispersion relations 
in the $xy$ plane and along the $z$ axis, 
the two kind groups of the OW with higher and lower energies appear, and 
the flat bands are not seen along the $(\pi \pi \pi)$-$(\pi \pi 0)$ direction. 
Actually, the inelastic neutron scattering experiments 
have started in YTiO$_3$. \cite{shamoto,ulrich}
The detailed comparison between the theoretical calculations 
and the experimental data can reveal 
nature of OW as well as that of OO in the $t_{2g}$ orbital systems.  

\begin{acknowledgments}
Author would like to thank S.~Maekawa, N.~Nagaosa, G.~Khaliullin, 
T.~Hatakeyama, and S.~Okamoto for their valuable discussions, 
and also thank Y.~Tokura, S.~Miyasaka, S.~Shamoto, and S.~Sugai for providing 
unpublished experimental data.
This work was supported by KAKENHI from MEXT, and 
KURATA foundation. 
Part of the numerical calculation has been performed by 
the supercomputing facilities in IMR, Tohoku University. 
\end{acknowledgments}

\begin{references}
\bibitem{tokura}
See, for a review, M.~Imada, A.~Fujimori, and Y.~Tokura, 
Rev. Mod. Phys. {\bf 70}, 1039 (1998).  
\bibitem{tokunaga}
Y.~Tokura and N.~Nagaosa, 
Science {\bf 288}, 462 (2000). 
%
\bibitem{cyrot}
M.~Cyrot, and C.~Lyon-Caen, 
Jour. Physique {\bf 36}, 253 (1975). 
\bibitem{komarov}
A.~G.~Komarov, and L.~I.~Korovin, and 
E.~K.~Kudinov, 
Sov. Phys. Solid State {\bf 17}, 1531 (1975). 
\bibitem{ishihara_eh}
S.~Ishihara, J.~Inoue, and S.~Maekawa, 
Phys. Rev. B {\bf 55}, 8280 (1997). 
\bibitem{saitoh}
E.~Saitoh, S.~Okamoto, K.~T.~Takahashi, K.~Tobe, K.~Yamamoto, 
T.~Kimura, S.~Ishihara, S.~Maekawa, and Y.~Tokura, 
Nature {\bf 410}, 180 (2001). 
\bibitem{gruninger}
M.~Gruninger, R.~Ruckamp, M.~Windt, 
P.~Reutler, C.~Zobel, T.~Lorenz, 
A.~Freimuth, and A.~Revcolevshi, 
Nature {\bf 418}, 39 (2002). 
\bibitem{saitoh_r}
E.~Saitoh, S.~Okamoto, K.~Tobe, K.~Yamamoto, 
T.~Kimura, S.~Ishihara, S.~Maekawa, and Y.~Tokura, 
Nature {\bf 418}, 40 (2002). 
%
\bibitem{nakao}
H.~Nakao, Y.~Wakabayashi, T.~Kiyama, 
Y.~Murakami, M.~v.~Zimmermann, 
J.~P.~Hill, D.~Gibbs, S.~Ishihara, Y.~Taguchi, and Y.~Tokura, 
Phys. Rev. B {\bf 66}, 184419 (2002). 
\bibitem{itoh}
M.~Itoh, M.~Tsuchiya, H.~Tanaka, and K.~Motoya, 
Jour. Phys. Soc. Jpn. {\bf 68}, 2783 (1999). 
\bibitem{ichikawa} 
H.~Ichikawa, J.~Akimitsu, M.~Nishi, and K.~Kakurai, 
Physica B {\bf 281$\&$282}, 482 (2000). 
\bibitem{maclean}
D.~A.~MacLean, H.-N.~Ng, and J.~E.~Greedan, 
Jour. Sol. Stat. Chem. {\bf 30}, 35 (1979).
%
\bibitem{ishihara_ti}
S.~Ishihara, T.~Hayakeyama, and S.~Maekawa, 
Phys. Rev. B {\bf 65}, 064442 (2002). 
%
\bibitem{mizokawa}
T.~Mizokawa, and A.~Fujimori, 
Phys. Rev. B {\bf 54}, 5368 (1996), and 
T.~Mizokawa, D.~I.~Khomskii, and G.~A.~Sawatzky, 
Phys. Rev. B {\bf 60}, 7309 (1999). 
\bibitem{sawada}
H.~Sawada, N.~Hamada, K.~Terakura, and T.~Asada, 
Phys. Rev. B {\bf 53}, 12742 (1996), and 
H.~Sawada, and K.~Terakura, 
Phys. Rev. B {\bf 58}, 6831 (1998). 
\bibitem{mochizuki}
M.~Mochizuki, and M.~Imada, 
Jour. Phys. Soc. Jpn. {\bf 69}, 1982 (2000), ibid 
{\bf 70}, 1777 (2001). 
%
\bibitem{ulrich}
C.~Ulrich, G.~Khaliullin, S.~Okamoto, M.~Reehuis, A.~Ivanov, 
H.~He, Y.~Taguchi, Y.~Tokura, and B.~Keimer, 
Phys. Rev. Lett. {\bf 89}, 167202 (2002). 
\bibitem{khaliullin}
G.~Khaliullin, and S.~Okamoto, 
Phys. Rev. Lett. {\bf 89}, 167201 (2002).
%
\bibitem{ren}
Y.~Ren, T.~T.~M.~Palstra, D.~I.~Khomskii, E.~Pellegrin, 
A.~A.~Nugroho, A.~A.~Menovski, and G.~A.~Sawatzky, 
Nature, {\bf 396}, 441 (1998). 
\bibitem{miyasaka}
S.~Miyasaka, Y.~Okimoto, and Y.~Tokura, 
Jour. Phys. Soc. Jpn. {\bf 71}, 2086 (2002). 
\bibitem{miyasaka2}
S.~Miyasaka, Y.~Okimoti, M.~Iwama, and Y.~Tokura, 
Phys. Rev. B {\bf 68}, 100406 (2003). 
%
\bibitem{bordet}
P.~Bordet, C.~Chaillout, M.~Marezio, Q.~Huang, 
A.~Santro, S.~-W.~Cheong, H.~Takagi, C.~S.~Oglesby, and 
B.~Batlogg, Jour. Sol. Stat. Chem. {\bf 106}, 253 (1993). 
\bibitem{kawano}
H.~Kawano, H.~Yoshizawa, and Y.~Ueda, 
Jour. Phys. Soc. Jpn. {\bf 63}, 2857 (1994). 
\bibitem{noguchi}
M.~Noguchi, A.~Nakazawa, S.~Oka, 
T.~Arima, Y.~Wakabayashi, H.~Nakao, and Y.~Murakami, 
Phys. Rev. B {\bf 62}, R9271 (2000). 
\bibitem{blake}
G.~R.~Blake, T.~T.~M.~Palstra, Y.~Ren, 
A.~A.~Nugroho, and A.~A.~Menovsky, 
Phys. Rev. Lett. {\bf 87}, 245501 (2001).  
\bibitem{ulrich_v}
C.~Ulrich, G.~Khaliullin, J.~Sirker, M.~Reehuis, M.~Ohl, 
S.~Miyasaka, Y.~Tokura, and B.~Keimer, 
(unpublished) cond-mat/0211589. 
%
\bibitem{khaliullin_v}
G.~Khaliullin, P.~Horsch, and A.~M.~Ol${\rm \acute e}$s, 
Phys. Rev. Lett. {\bf 86}, 3879 (2001), and 
J.~Sirker, and G.~Khaliullin, 
Phys. Rev. B {\bf 67}, 100408 (2003). 
\bibitem{motome}
Y.~Motome, H.~Seo, Z.~Fang, and N.~Nagaosa, 
Phys. Rev. Lett. {\bf 90}, 146602 (2003). 
\bibitem{fang}
Z.~Fang, N.~Nagaosa, and K.~Terakura,  
Phys. Rev. B {\bf 67}, 035101 (2003). 
\bibitem{silva}
T.~H.~De Silva, A.~Joshi, M.~Ma, and F.~C.~Zhang, 
(unpublished) cond-mat/03024989. 
%
%
\bibitem{gellmann}
M.~Gell-Mann, and Y.~Ne'eman. 
{\it The eightfold way}, (Benjamin, New York, 1964).  
%
%
\bibitem{kugel}
K.~I.~Kugel, and D.~I.~Khomskii, 
Sov. Phys. Solid, State, {\bf 17}, 285 (1975).
\bibitem{kikoin}
K.~Kikoin, O.~Entin-Wohlman, 
V.~Fleurov, and A.~Aharony, 
Phys. Rev. B {\bf 67}, 214418 (2003). 
%
\bibitem{itoh_m}
M.~Itoh, H.~Adachi, H.~Nakao, Y.~Murakami, Y.~Taguchi, Y.~Tokura, K.~Kato, E.~Nishibori, 
M.~Takata, M.~Sakata, H.~Miyagawa, S.~Nanao, H.~Maruyama, E.~Arawaka, and K.~Namikawa, 
Meeting Abstract of the Physicsl Society of Japan, 
2001 Autumn Meeting, 
{\bf 56}, Part 3, 341 (2001). 
%
\bibitem{janssen}
D.~Janssen, R.~V.~Jolos, and F.~D${\rm o}$nau, 
Nucl. Phys. A {\bf 224}, 93 (1974). 
\bibitem{klein}
A.~Klein, and E.~R.~Marshalek, 
Rev. Mod. Phys. {\bf 63}, 375 (1991). 
%
\bibitem{inoue}
J.~Inoue, S.~Okamoto, S.~Ishihara, W.~Koshibae, Y.~Kawamura, and S.~Maekawa, 
Physica B  {\bf 237-238}, 51 (1997).  
\bibitem{okamoto_ow}
S.~Okamoto, S.~Ishihara, and S.~Maekawa,   
Phys. Rev. B {\bf 66}, 104435 (2001).  
\bibitem{allen}
P.~B.~Allen, and V.~Pereveinos, 
Phys. Rev. Lett. {\bf 83}, 4828 (1999), 
and 
V.~Pereveinos, and P.~B.~Allen, 
Phys. Rev. B {\bf 64}, 085118 (2001). 
%
\bibitem{miyasaka3}
S.~Miyasaka and Y.~Tokura (unpublished). 
\bibitem{shamoto}
S.~Shamoto, F.~Iga, M.~Tsubota, T.~Kajitani, 
ISIS Experimental Report, RB Num.13662, 
Date of Report 03/03/03, MARI.
\end{references}
\end{document}